\documentstyle[multicol,prl,aps,epsf]{revtex}
\begin{document}
\draft
\preprint{}
\title{Edge Singularity for the Optically Induced Kondo 
Effect in a Quantum Dot}
\author{T.\ Fujii,$^1$ A.\ Furusaki,$^2$ N.\ Kawakami$^1$ 
and M.\ Sigrist$^2$} 
\address{
$^1$Department of Applied Physics,
Osaka University, Suita, Osaka 565, Japan \\
$^2$Yukawa Institute for Theoretical Physics, 
Kyoto University, Kyoto 606-8502, Japan} 
\date{\today}
\maketitle
%%%%%%%%%%%%%%%                                          %%%%%%%%%%%%%%%%
%%%%%%%%%%%%%%%                ABSTRACT                  %%%%%%%%%%%%%%%%
%%%%%%%%%%%%%%%                                          %%%%%%%%%%%%%%%%
\begin{abstract}
We analyze the optically induced Kondo effect in the
absorption spectrum for a quantum dot 
with an even number of electrons, for which the Kondo effect 
does not occur in the ground state. 
The Kondo exchange couplings generated for 
photo-excited states can be either antiferromagnetic or
ferromagnetic, which may result in different types of the
edge singularities, depending on the microscopic parameters of the dot.
We discuss critical properties
in the spectrum by using low-energy   effective field theory.
\end{abstract}
%%%%%%%%%%%%%%%%%%%%%%%%%%%%%%%%%%%%%%%%%%%%%%%%%%%%%%%%%%%%%%%%%%

\begin{multicols}{2}

Recently quantum dots have attracted much attention.
They provide useful systems to study the effects of
local electron correlations, 
for which the microscopic parameters can be varied
systematically. For instance, the Kondo effect found in 
resonant tunneling phenomena has been 
intensively studied theoretically
\cite{Ng,Glazman,Kawabata,Hershfield,Meir}
and experimentally\cite{Goldhaber,Cronenwett}.
Besides such transport experiments, optical probes
are also useful to explore many-body effects in quantum dots.
Although optical experiments have revealed a number of
interesting properties, they have been mainly concerned with
the properties of charge excitations.
Magnetic correlations in the optical spectrum for
quantum dots have not been studied systematically thus far.
Recently, Kikoin {\it et al.} discussed how the spin degrees of
freedom can affect the optical line shape for a quantum dot
with an even number of electrons, for which the Kondo-type 
correlation is developed in photo-excited 
states\cite{Kikoin}.
In a similar context, it was demonstrated that 
such Kondo-type correlations may appear in
photoemission experiments, resulting in anomalous 
edge-singularity properties in the spectrum\cite{fujii3}. 
It was also pointed out that 
the non linear optical response is affected by 
these spin correlations\cite{shahba}. 
These Kondo-type effects in dynamical quantities such as the optical
conductivity and the photoemission spectrum, 
may be referred to as  the {\it dynamically induced Kondo effect}. 

In this paper we study 
the dynamically induced Kondo effect in the optical absorption 
spectrum of a quantum dot having two levels with no spin moment in the
ground state.
We show that, in the optically excited state where each level is
occupied by one electron, the effective spin moment in each level has
a Kondo exchange coupling to the leads, which can be either
antiferromagnetic or ferromagnetic. 
We use an effective Kondo model to analyze low-energy critical
properties in the optical absorption spectrum
around the particle-hole excitation energy of the dot. 
We find that the effective Kondo coupling leads to 
strongly field-dependent edge singularity. 

We start with a two-level quantum dot which has two leads
attached.
This system is described by the impurity Anderson model
involving two orbitals with the Hamiltonian\cite{Sakai,Izumida,Yeyati},
%%%%%%%%%%%%%%%%%%%%%%%%%%%%%%%%%%%%%%%%%%%%%%%%%%%%%%%
\begin{eqnarray}
 H&=& \sum_{bk\sigma} 
     {\epsilon}_k c^{\dagger}_{bk\sigma} c_{bk\sigma}
      +\sum_{\alpha \sigma} {\varepsilon}_\alpha n_{\alpha \sigma}
       +U\sum_\alpha n_{\alpha \uparrow} n_{\alpha \downarrow}
             \nonumber\\
 & &
      +U^{'}\sum_{\sigma {\sigma}^{'}}n_{1 \sigma} n_{2 {\sigma}^{'}}
      + \sum_{b\alpha k\sigma}
          \left(V_{b\alpha} c^{\dagger}_{bk\sigma} d_{\alpha \sigma}
           +{\rm h.c.}\right),
\label{Hamiltonian}
\end{eqnarray}
%%%%%%%%%%%%%%%%%%%%%%%%%%%%%%%%%%%%%%%%%%%%%%%%%%%%%%%%
where $c_{lk\sigma}$ ($c_{rk\sigma}$) annihilates an electron 
in the left (right) leads,
$d_{\alpha \sigma}$ annihilates an electron in the orbital
$\alpha$ in the dot ($\alpha=1,2$), and
$n_{\alpha\sigma}=d^\dagger_{\alpha\sigma}d_{\alpha\sigma}$.
Both intra- and inter-orbital Coulomb repulsion in the dot are
included.
The last term with the coupling $V_{b\alpha}$ describes the tunneling
between the dot and the leads and will be denoted as $H_T$
($ H=H_0 + H_T $). 
We will consider the optical transition from the state where only the
lower ($\alpha=1$) orbital is occupied by two electrons to a
state where each orbital is occupied by one electron. 
The spin of each orbital in the excited state is subject to the 
Kondo exchange interaction with the electrons in the leads.
For the antiferromagnetic coupling of our main interest, the presence
of two leads allows the screening of both spins by two 
{\em orthogonal} channels of the lead electrons.
The effective decoupling of screening channels becomes immediately
clear for the case where the two Kondo states have very different
Kondo temperatures and one spin is completely screened at a much
higher energy than the other.
The residual Kondo coupling leads then
to the Kondo effect for the second spin. 
Therefore, without losing generality we may use a simple illustrative
model where the two channels coupled to the two spins are independent
and orthogonal to each other in an obvious form already in the bare
Hamiltonian.
We assume that the quantum dot including leads has the reflection
symmetry with respect to a mirror plane in the center, so that we may
classify states by parity.
Note that this is also compatible with the requirement of an optical
(dipole) transition.
We thus assume that the hybridizations $V_{b\alpha}$ is
either symmetric or antisymmetric:
%%%%%%%%%%%%%%%%%%%%%%%%%%%%%%%%%%%%%%%%%%%%%%%%%%%%%%%
$
 V_{l1}=V_{r1} \equiv  V_1/(2\sqrt{2}) ,{\ }
 V_{l2}=-V_{r2} \equiv V_2/(2\sqrt{2}).
$
%%%%%%%%%%%%%%%%%%%%%%%%%%%%%%%%%%%%%%%%%%%%%%%%%%%%%%%

We are interested in the ground-state configuration where two
electrons occupy the lower level ($\alpha=1$) in the dot.
This is realized as a lowest energy state when the conditions
%%%%%%%%%%%%%%%%%%%%%%%%%%%%%%%%%%%%%%%%%%%%%%%%%%%%%%%
\begin{equation}
 \varepsilon_2 >\varepsilon_1 + U-U', \quad 
  \varepsilon_1 <-U, \quad
   \varepsilon_2 >-2U'
\label{condition}
\end{equation}
%%%%%%%%%%%%%%%%%%%%%%%%%%%%%%%%%%%%%%%%%%%%%%%%%%%%%%%
are satisfied, and when the tunneling is very weak
$\Gamma_\alpha\ll\Delta_0$.
Here $\Delta_0 =\varepsilon_2 -\varepsilon_1 -U+U'$ is 
the bare particle-hole excitation energy in the dot,
and $\Gamma_\alpha=\pi|V_\alpha|^2/D$ is the level width with $D$
being the band width in the leads.
Since there is no effective spin moment in the dot, the Kondo effect
does not occur in the ground state.
As demonstrated by Kikoin {\it et al.}\cite{Kikoin,fujii3}, however,
the Kondo effect shows up in optical absorption spectra, whose
critical behavior will be studied in detail below. 

\begin{figure}[htb]
%\hspace*{9mm}
\begin{center}
\epsfxsize=70mm
%\epsfbox{dot1.eps}
\epsfbox{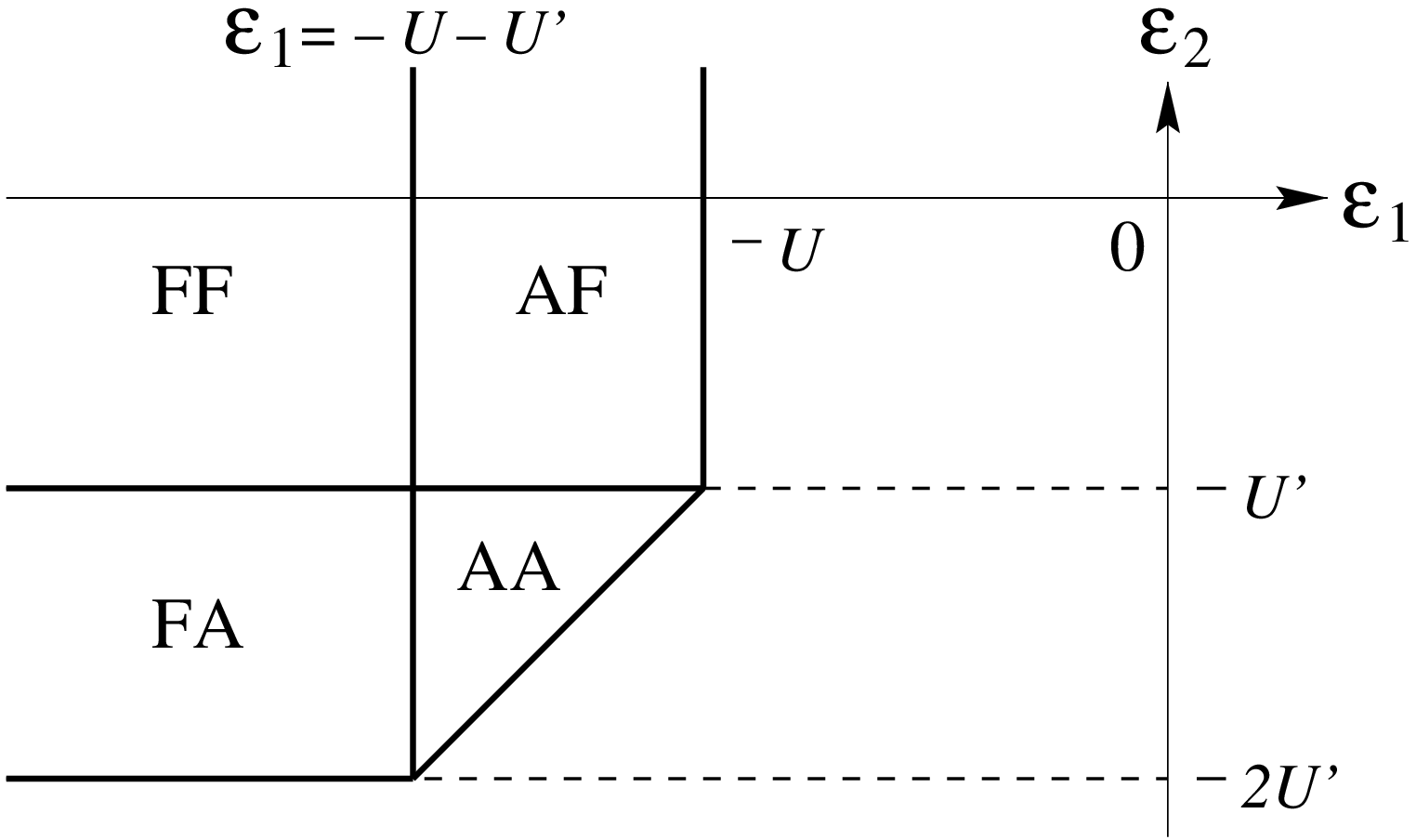}
{\narrowtext
\caption{Energy-level diagram specified by the two Kondo exchange 
couplings  $J_{K \alpha}$ which can be either positive or negative
within the conditions (\protect{\ref{condition}}).
For example, in the region denoted as AF, $J_{K1}$ is
antiferromagnetic and $J_{K2}$ is ferromagnetic.
}
\label{fig:diagram}
}
\end{center}
\end{figure}

The optical absorption spectrum of the dot is given by 
%%%%%%%%%%%%%%%%%%%%%%%%%%%%%%%%%%%%%%%%%%%%%%%%%%%%%%%
\begin{equation}
 I(\omega)
  =-2|M|^2 {\rm Im}\,G^{>}(\omega+i\delta),
\label{I(omega)}
\end{equation}
%%%%%%%%%%%%%%%%%%%%%%%%%%%%%%%%%%%%%%%%%%%%%%%%%%%%%%%
where $\delta$ is positive infinitesimal and $M$ represents 
the optical transition matrix element between the two quantum dot
states. 
The dynamical correlation function, $G^{>}$, is
expressed as
%%%%%%%%%%%%%%%%%%%%%%%%%%%%%%%%%%%%%%%%%%%%%%%%%%%%%%%
\begin{equation}
 G^{>}(\omega)
  = \frac{1}{\omega -\Delta_0} + 
     \frac{1}{(\omega -\Delta_0)^2}
      \langle 0|{\cal O} T(\omega) {\cal O}^{\dagger}|0\rangle.
\label{G^>}
\end{equation}
%%%%%%%%%%%%%%%%%%%%%%%%%%%%%%%%%%%%%%%%%%%%%%%%%%%%%%%
With the conditions (\ref{condition}), the ground state
$|0\rangle$ is given by
$|0\rangle=d_{1\uparrow}^{\dagger}
           d_{1\downarrow}^{\dagger} |F\rangle$,
where $|F\rangle$ is the filled Fermi sea.
The optical transition is caused by the operator
${\cal O}^{\dagger} = (1/\sqrt{2})
  \sum_\sigma d_{2\sigma}^{\dagger}d_{1\sigma}$. 
Up to second order in $H_T$ the $T$-matrix, defined by
$
T(\omega)=H_T+H_T(\omega-H_0)^{-1}H_T+\cdots,
$
is reduced to the effective Hamiltonian for the excited state
${\cal O}^\dagger|0\rangle$,
%%%%%%%%%%%%%%%%%%%%%%%%%%%%%%%%%%%%%%%%%%%%%%%%%%%%%%%
\begin{eqnarray*}
 H_{\rm eff}
  &=&\sum_\alpha
   \left(
    -\frac{ \Gamma_\alpha }{ \pi } 
      \ln{\frac{D}{-\epsilon_\alpha - U}}
       +V_{\alpha P} \sum_{\sigma}
        \psi_{\alpha\sigma}^{\dagger}(0)\psi_{\alpha\sigma}(0)
         \right. \cr
  & & {\quad \quad}+4\pi J_{K \alpha} \sum_{\sigma\sigma^{'}}
          \psi_{\alpha\sigma}^{\dagger}(0)
           \frac{\sigma_{\sigma\sigma^{'}}^a}{2}
            \psi_{\alpha\sigma^{'}} (0)
           S_{\alpha}^a 
   \Biggr).
%\label{H_eff}
\end{eqnarray*}
%%%%%%%%%%%%%%%%%%%%%%%%%%%%%%%%%%%%%%%%%%%%%%%%%%%%%%%
Here $\sigma^a$ is the Pauli matrix, and 
$S^a_{\alpha}=
(1/2)d^\dagger_{\alpha\sigma}
\sigma^a_{\sigma\sigma'}d_{\alpha\sigma'}$
are effective impurity spins formed in the two levels in the excited
state ${\cal O}^\dagger|0\rangle$.
The impurity spins are coupled to electrons in the leads, 
$
%%\begin{eqnarray*}
 \psi_{1\sigma}(0)=
  (1/{\sqrt2})\sum_{k}(c_{lk\sigma}+c_{rk\sigma}),{\ } 
 \psi_{2\sigma}(0)=
  (1/{\sqrt2})\sum_{k}(c_{lk\sigma}-c_{rk\sigma}),
%%\end{eqnarray*} 
$
via two kinds of Kondo exchange interactions
\begin{equation}
\displaystyle
J_{K \alpha}
  =
   |V_\alpha|^2
    \left(
     -\frac{1}{\varepsilon_\alpha +U'}
     +\frac{1}{\varepsilon_\alpha +U'+U}
    \right).
\label{J_K}
\end{equation}
Note that the Kondo couplings can be either antiferromagnetic or
ferromagnetic because the energy of intermediate states 
measured from the optically excited states,
$-\varepsilon_\alpha -U'$ and 
$\varepsilon_\alpha +U'+U$, are not necessarily positive.
The Kondo couplings cause final-state interactions
as in the Fermi-edge singularity\cite{Schotte,Ludwig2}, giving rise to
interesting features in the absorption spectrum as we will discuss
below.
The parameter region of our interest is shown in
Fig.~\ref{fig:diagram}.
We note that an antiferromagnetic Kondo coupling is not generated 
without the inter-orbital Coulomb interaction $U'$, in which case
the Kondo screening does not occur in the optically excited state.

%%%%%%%%%%%%%%%%%   FIGURE 1   %%%%%%%%%%%%%%%%%%%%%%%
%%%%%%%%%%%%%%%%%%%%%%%%%%%%%%%%%%%%%%%%%%%%%%%%%%%%%

We will analyze the critical behavior in the optical absorption
spectrum in detail for the following two cases:
(i) $J_{K1}>0$ and $J_{K2}>0$
and (ii) $J_{K1}<0$ and $J_{K2}>0$.
We will first ignore the potential scattering term $V_{\alpha P}$
and concentrate on the dynamical properties of spin degree of
freedom. 
The potential scattering affects only the charge sector and its effect 
will be discussed shortly.
The Kondo effect in $H_{\rm eff}$ is none but two decoupled
single-channel Kondo problems, whose low-energy physics can be studied
in the standard manner using a one-dimensional (1D) model for
electrons in the leads \cite{Affleck,Ludwig}.
We thus introduce two left-going electron modes with linear dispersion
$\psi_{\alpha L\sigma}(x)$ defined on a full line ($-\infty<x<\infty$),
which satisfy $\psi_{\alpha L\sigma}(0)=\psi_{\alpha\sigma}(0)$.
We then use nonabelian bosonization methods to rewrite the Hamiltonian.
In this representation the charge and spin sectors are decoupled and
the spin part of the kinetic energy of 1D electrons reads
%%%%%%%%%%%%%%%%%%%%%%%%%%%%%%%%%%%%%%%%%%%%%%%%%%%%%%%
\begin{equation}
{\cal H}_0=
\frac{v_F}{2\pi} \sum_\alpha \int^{\infty}_{-\infty}dx
       \frac{1}{k+2} :J_{\alpha L}^a (x) J_{\alpha L}^a (x):,
\end{equation}
where $v_F$ is the Fermi velocity and the level $k=1$ SU(2) currents
are defined by 
%%%%%%%%%%%%%%%%%%%%%%%%%%%%%%%%%%%%%%%%%%%%%%%%%%%%%%%
\begin{eqnarray}
 J_{\alpha L}^a (x)=\sum_{\sigma \sigma^{'}}
  :\psi_{\alpha L\sigma}^{\dagger} (x)
       \frac{\sigma^{a}_{\sigma \sigma^{'}}}{2}
        \psi_{\alpha L\sigma} (x):. 
\label{SU(2)currents}
\end{eqnarray}
%%%%%%%%%%%%%%%%%%%%%%%%%%%%%%%%%%%%%%%%%%%%%%%%%%%%%%%
The Kondo interactions in $H_{\rm eff}$ is written
\begin{equation}
{\cal H}_K=
\sum_\alpha \lambda_{K \alpha}J_{\alpha L}^a (0)S_{\alpha}^a,
\label{H_K}
\end{equation}
%%%%%%%%%%%%%%%%%%%%%%%%%%%%%%%%%%%%%%%%%%%%%%%%%%%%%%%
where  $\lambda_{K \alpha}=4\pi J_{K \alpha}/v_F$.
Finally the optical absorption spectrum $I(\omega)$ is given by the
Fourier transform of
%%%%%%%%%%%%%%%%%%%%%%%%%%%%%%%%%%%%%%%%%%%%%%%%%%%%%%%
\begin{equation}
 G(t)=\langle0|{\cal O} e^{i {\cal H}_0 t}
      e^{-i ({\cal H}_0+{\cal H}_K) t}
      {\cal O}^{\dagger}|0\rangle
      e^{-i \Delta t},
\label{G(t)}
\end{equation}
%%%%%%%%%%%%%%%%%%%%%%%%%%%%%%%%%%%%%%%%%%%%%%%%%%%%%%%
where
\begin{equation}
\displaystyle \Delta =\Delta_0 
 -\sum_\alpha \frac{ \Gamma_\alpha }{\pi} 
   \ln{\frac{D}{-\epsilon_\alpha -U}}
\end{equation}
is the renormalized particle-hole excitation energy.

{\em (i) region AA} with $J_{K1}>0$ and $J_{K2}>0$.
The weak antiferromagnetic Kondo couplings are renormalized to the
strong-coupling fixed point, where
$
\lambda_{K \alpha} 
 \rightarrow \lambda^{*}_{K \alpha}
  =2/(k+2)
$ with $k=1$. 
%%%%%%%%%%%%%%%%%%%%%%%%%%%%%%%%%%%%%%%%%%%%%%%%%%%%%%%
Following Refs.~\onlinecite{Affleck,Ludwig}, we introduce the
unitary operators $U_\alpha$ transforming the SU(2) currents as
$
U_\alpha J_{\alpha L}^a (x) U_\alpha^{\dagger}=
J_{\alpha L}^a (x) + 2\pi S_{\alpha}^a \delta (x) 
\equiv {\cal J}_{\alpha L}^a (x)
$. 
The unitary operators $ U_1 $ and $ U_2$ commute with each other
because the Hamiltonian is decoupled into two sectors.
The merit of using $U$ is that it allows us to rewrite the Hamiltonian
as 
%%%%%%%%%%%%%%%%%%%%%%%%%%%%%%%%%%%%%%%%%%%%%%%%%%%%%%%
\begin{equation}
 U {\cal H}_0 U^{\dagger} = {\cal H}_0 + {\cal H}_K
\label{UH_0U}
\end{equation}
%%%%%%%%%%%%%%%%%%%%%%%%%%%%%%%%%%%%%%%%%%%%%%%%%%%%%%%
at $\lambda_{K \alpha}=\lambda^*_{K \alpha}$.
The long-time behavior of $G(t)$ is then obtained from the correlation 
function of $U_\alpha$, 
%%%%%%%%%%%%%%%%%%%%%%%%%%%%%%%%%%%%%%%%%%%%%%%%%%%%%%%
\begin{eqnarray}
 G(t)\sim
  \langle U_1 (t) U_1^{\dagger}(0)\rangle
  \langle U_2 (t) U_2^{\dagger}(0)\rangle
  e^{-i \Delta t}
\label{G(t) 2}
\end{eqnarray}
%%%%%%%%%%%%%%%%%%%%%%%%%%%%%%%%%%%%%%%%%%%%%%%%%%%%%%%
where 
%%%%%%%%%%%%%%%%%%%%%%%%%%%%%%%%%%%%%%%%%%%%%%%%%%%%%%%
$
 U_\alpha (t) 
  = e^{i{\cal H}_{0}t} 
   U_\alpha
    e^{-i{\cal H}_{0}t}
$
%%%%%%%%%%%%%%%%%%%%%%%%%%%%%%%%%%%%%%%%%%%%%%%%%
can be regarded as a boundary condition changing operator
\cite{Cardy2,Ludwig2} and the brackets represents the average at the
strong-coupling fixed point.
We may rewrite (\ref{UH_0U}) by 
using the operator product expansion (OPE) 
in terms of the new currents ${\cal J}^a_{\alpha L}$:
%%%%%%%%%%%%%%%%%%%%%%%%%%%%%%%%%%%%%%%%%%%%%%%%%%%%%%%
\begin{equation}
 {\cal J}^a_{\alpha L} (z) U^{\dagger}_\alpha (w)
  = \frac{-S^a_\alpha}{z-w} 
   U^{\dagger}_\alpha (w)
    + {\rm reg.} \cdots,
\label{OPE}
\end{equation}
%%%%%%%%%%%%%%%%%%%%%%%%%%%%%%%%%%%%%%%%%%%%%%%%%%%%%%%
which just corresponds to the OPE for 
the highest weight of  SU(2) current algebra. 
Accordingly, the correlation function for
$U^{\dagger}_\alpha (w)$ satisfies the so-called 
Knizhnik-Zamolodchikov equation\cite{Knizhnik},
%%%%%%%%%%%%%%%%%%%%%%%%%%%%%%%%%%%%%%%%%%%%%%%%%%%%%%%
\begin{equation}
 \left(
  \frac{\partial}{\partial z_{i}}+
   \frac{2}{k+2} \sum_{l \neq i}
    \frac{{\hat{S}}^a_{i \alpha} 
     \otimes {\hat{S}}^a_{l \alpha}}
     {z_i -z_l}
 \right)
     \langle U_\alpha (z_i) U^{\dagger}_\alpha (z_l)\rangle
       =0,
\label{KZeq}
\end{equation}
%%%%%%%%%%%%%%%%%%%%%%%%%%%%%%%%%%%%%%%%%%%%%%%%%%%%%%%
where the operator ${\hat{S}}^a_{i \alpha}$ associated with 
$U_\alpha$ acts from the right like 
$ {\hat S}^a_{i \alpha} \cdot U_\alpha (z_i)
= U_\alpha (z_i)S^a_\alpha$
whereas  ${\hat{S}}^a_{l \alpha}$ associated with
$U^{\dagger}_\alpha$ acts from the left.
Since the impurity spin $S^a_\alpha$ 
satisfies $\sum_a S^a_\alpha S^a_\alpha =S(S+1)$,
the solution of Eq.~(\ref{KZeq}) is
%%%%%%%%%%%%%%%%%%%%%%%%%%%%%%%%%%%%%%%%%%%%%%%%%%%%%%%
\begin{equation}
 \langle U_\alpha (t) U^{\dagger}_\alpha (0)\rangle
  = \frac{1}{t^{2\Delta_s}}
\label{power-law decay}
\end{equation}
%%%%%%%%%%%%%%%%%%%%%%%%%%%%%%%%%%%%%%%%%%%%%%%%%%%%%%%
with the boundary dimension,
%%%%%%%%%%%%%%%%%%%%%%%%%%%%%%%%%%%%%%%%%%%%%%%%%%%%%%%
\begin{equation}
 \Delta_s =\frac{S(S+1)}{k+2}.
\label{Delta_s}
\end{equation}
%%%%%%%%%%%%%%%%%%%%%%%%%%%%%%%%%%%%%%%%%%%%%%%%%%%%%%%
For the present model ($S=1/2, k=1$), $\Delta_s=1/4$ \cite{Ludwig2}
for each channel $\alpha$.

The Kondo temperature can be different for two channels\cite{Izumida},
depending on the parameters
$\varepsilon_\alpha$, $U$, and $\Gamma_\alpha$.
In general we can expect two critical regions to appear in the
absorption spectrum: 
%%%%%%%%%%%%%%%%%%%%%%%%%%%%%%%%%%%%%%%%%%%%%%%%%%%%%%%
\begin{equation}
 I(\omega) = 
  \left\{
    \begin{array}{rl}
       (\omega -\Delta)^{4\Delta_s -1}, 
      &{\ } 0< \omega-\Delta \ll T_{K1}, \cr
       (\omega -\Delta)^{2\Delta_s -1},
      &{\ }  T_{K1} \ll \omega -\Delta \ll T_{K2}.
    \end{array}
  \right.
\label{two regimes}
\end{equation}
%%%%%%%%%%%%%%%%%%%%%%%%%%%%%%%%%%%%%%%%%%%%%%%%%%%%%%%
%
%As typical values for a small quantum dot, let us set the parameters
%as 
%$\varepsilon_1= -1.5 {\rm meV}$,
%$\varepsilon_2= -1.1 {\rm meV}$,
%$D=5{\rm meV}$,
%$\Gamma_1=0.075{\rm meV}$,
%$\Gamma_2=0.2{\rm meV}$, and
%$U=U^{'}=2{\rm meV}$.
%We then find two Kondo temperatures,
%$T_{K1} =5 \times 10^{-9} {\rm meV}$ and 
%$T_{K2} =1.12 {\rm meV}$, which are quite different in magnitude,
%$T_{K1}\ll T_{K2}$.
%Because the energy fluctuation 
%due to the life-time of the exciton 
%in ordinary semiconductors 
%is about $4.1 \times 10^{-6}{\rm meV}$, 
%the critical region, 
%$\omega -\Delta\ll T_{K1}$, may  not be
%observed in actual dot systems.
% 
For typical values of small quantum dots we may encounter a situation
where, say, $T_{K1}\ll T_{K2}$ because the deeper level may have a
smaller $\Gamma_1$.
In this case the impurity spin $S_1$ remains essentially free whereas
the spin $S_2$ is screened, and
the Kondo effect with the critical 
behavior $(\omega-\Delta)^{-1/2}$ may be observed in the low-energy
regime of the absorption spectrum.
A twist to the argument occurs when we introduce 
Hund's-rule coupling $J_H S^a_1 S^a_2$ between the spins of the two
orbitals \cite{Tarucha}.
In this case the coupling will 
generate a logarithmic correction to the $(\omega-\Delta)^{-1/2}$
behavior in $T_{K1}\ll T \ll  T_{K2}$, 
as we will show later [Eq.~(\ref{log-correction})].

So far we have exploited SU(2) current algebra to
discuss low-energy properties of the absorption spectrum.
Although this analysis cannot be used when the magnetic 
field is applied, we can still evaluate 
the boundary dimension  $\Delta_s$
as a function of magnetic fields, by
applying the finite-size scaling to 
the exact solution of the Kondo model\cite{Andrei,Tsvelik}. 
Namely, the scaling dimension $\Delta_s$ is
determined by analyzing the excitation spectrum in the 
presence of the Kondo impurity.  
Following the analysis in \onlinecite{fujii3}, we obtain
%%%%%%%%%%%%%%%%%%%%%%%%%%%%%%%%%%%%%%%%%%%%%%%%%%%%%%%
\begin{equation}
 \Delta_s = n_{{\rm imp}}^2,
  \qquad
 n_{\rm imp} = \int_{-\infty}^{\Lambda_0}
         \sigma_{\rm imp} (\Lambda) d\Lambda .
\label{Delta_s in H}
\end{equation}
%%%%%%%%%%%%%%%%%%%%%%%%%%%%%%%%%%%%%%%%%%%%%%%%%%%%%%%
Here $\Lambda_0=-1/J+(1/\pi)\ln(\sqrt{2}T_H/H)$ and
$\sigma_{\rm imp}$ is determined by 
the integral equation for the exact solution 
of the Kondo model\cite{Andrei,Tsvelik}, 
%%%%%%%%%%%%%%%%%%%%%%%%%%%%%%%%%%%%%%%%%%%%%%%%%%%%%%%
\begin{eqnarray}
 \sigma_{\rm imp} (\Lambda) 
   &=&
   \frac{1}{2\pi[(\Lambda+1/J)^2+1/4]} \nonumber\\
   &&
        -\int_{-\infty}^{\Lambda_0}
        \frac{\sigma_{\rm imp} (\Lambda^{'})}
             {\pi[(\Lambda-\Lambda^{'})^2+1]} 
        d\Lambda^{'}. 
\nonumber
%\label{integral equation}
\end{eqnarray}
%%%%%%%%%%%%%%%%%%%%%%%%%%%%%%%%%%%%%%%%%%%%%%%%%%%%%%%
The critical exponent thus obtained for 
the absorption spectrum is shown in Fig.~2 as a 
function of the magnetic field. 
%%%%%%%%%%%%%%%%%  FIGURE  2  %%%%%%%%%%%%%%%%%%%%%%%%%%
\begin{figure}[htb]
\epsfxsize=72mm
\epsfbox{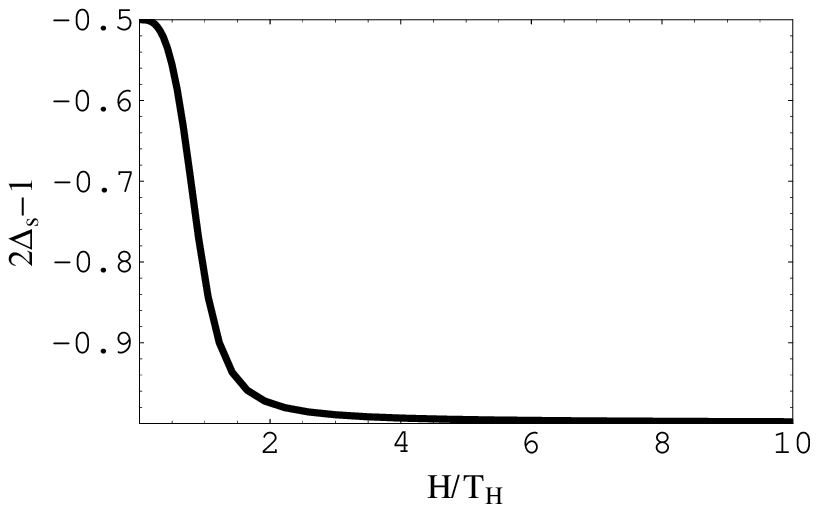}
{\narrowtext
\caption{The magnetic-field dependence of the critical exponent 
$2\Delta_s -1$. $T_H$ is the Kondo temperature 
in magnetic fields defined in \protect\onlinecite{Andrei,Tsvelik}}
\label{fig:2}
}
\end{figure}
%%%%%%%%%%%%%%%%%%%%%%%%%%%%%%%%%%%%%%%%%%%%%%%%%%%%%%%
At $H=0$, we recover $\Delta_s=1/4$,  
obtained above by current algebra techniques. 
When the potential scattering terms $V_{\alpha P}$ in
$H_{\rm eff}$ neglected so far
is incorporated\cite{Schotte,Ludwig2}, it will lead to an additional
power-law factor $t^{-2\Delta_c}$ to $G(t)$, where the scaling
dimension $\Delta_c$ for the charge sector is determined by the phase
shift due to the potential scattering.
The divergence singularity in the
spectrum is weakened by the additional exponent $2\Delta_c$.
Since $\Delta_c$ has little dependence on the magnetic field,
the field-dependence of the critical exponent
is solely determined by the dynamically induced Kondo effect,
as shown in Fig.~\ref{fig:2}.

{\em (ii) region FA} with $J_{K1}<0$ and $J_{K2}>0$.
We now turn to the case where only one of the Kondo
couplings is antiferromagnetic.
In this case, the impurity spin $S_1$ is decoupled from the lead
electrons at the fixed point because the ferromagnetic Kondo coupling
is marginally irrelevant,
whereas the impurity spin $S_2$ is completely screened due to
the Kondo effect. Note that a somewhat similar situation occurs in the
intermediate temperature range $T_{K1}\ll T \ll  T_{K2}$ for the case
discussed above, where one spin is screened while the other is still
basically free.  
With the inclusion of the Hund's-rule coupling between $S_1$ and $S_2$,
the two spins may be effectively combined to form a spin triplet
$S=1$, which is then screened by only one channel ($\alpha=2$) of
electrons, the so-called underscreened Kondo system is realized
\cite{Nozieres}.
It is known that for the fixed point of this system
the decoupled impurity $S_1$ weakly interacts with the renormalized
current ${\cal J}^a_{2L}$ via the ferromagnetic Kondo interaction, 
$\lambda_3 {\cal J}^a_{2L}(t)S^a_1$.
Although the coupling is marginally irrelevant\cite{Nozieres,Kusunose} 
and leads to $\langle U_1(t)U_1^\dagger(0)\rangle\sim{\rm const.}$, it
still affects the long-time
behavior of $\langle U_2(t)U_2^\dagger(0)\rangle$.
We can explicitly evaluate it around the fixed point, e.g., by
solving Callan-Symanzik equation in the renormalization group, 
%%%%%%%%%%%%%%%%%%%%%%%%%%%%%%%%%%%%%%%%%%%%%%%%%%%%%%%
\begin{eqnarray}
 \left(
  \frac{\partial}{\partial \ln t} +2\gamma_2
  - \beta_2 \frac{\partial}{\partial \lambda_2}
  - \beta_3 \frac{\partial}{\partial \lambda_3} 
 \right)
   \langle U_2(t)U_2^{\dagger}(0)\rangle=0.
\nonumber
\end{eqnarray}
%%%%%%%%%%%%%%%%%%%%%%%%%%%%%%%%%%%%%%%%%%%%%%%%%%%%%%%
Here $\beta_2$ is the beta function for $\lambda_2 U_2$ whereas 
$\beta_3$ is that for the effective ferromagnetic coupling
$\lambda_3 {\cal J}^a_{2L}(t)S^a_1$ 
caused by the Hund's-rule coupling.
The corresponding scaling equations read
%%%%%%%%%%%%%%%%%%%%%%%%%%%%%%%%%%%%%%%%%%%%%%%%%%%%%%%
\begin{eqnarray}
 \beta_2&=&\frac{d\lambda_2}{d\ln t}
        =\frac{3}{4}\lambda_2
         +\frac{1}{4}\lambda_2 \lambda_3, \cr
 \beta_3&=&\frac{d\lambda_3}{d\ln t}
        =\lambda^2_3
         +\frac{1}{4}\lambda_2^2, 
\label{eq24}
\end{eqnarray}
%%%%%%%%%%%%%%%%%%%%%%%%%%%%%%%%%%%%%%%%%%%%%%%%%%%%%%%
where
$
 \gamma_2 \simeq 1- \frac{\partial \beta_2}
                         {\partial \lambda_2}. 
$
%%%%%%%%%%%%%%%%%%%%%%%%%%%%%%%%%%%%%%%%%%%%%%%%%%%%%%%
By solving these equations, we have the leading-order 
contribution
%%%%%%%%%%%%%%%%%%%%%%%%%%%%%%%%%%%%%%%%%%%%%%%%%%%%%%%
\begin{eqnarray}
 \langle U_2(t)U_2^{\dagger}(0)\rangle
 \propto
 \sqrt{\frac{\ln t}{t}}.
\label{eq25}
\end{eqnarray}
%%%%%%%%%%%%%%%%%%%%%%%%%%%%%%%%%%%%%%%%%%%%%%%%%%%%%%%
This leads to the
low-energy behavior of
the absorption spectra, 
%%%%%%%%%%%%%%%%%%%%%%%%%%%%%%%%%%%%%%%%%%%%%%%%%%%%%%%
\begin{eqnarray}
 I(\omega) \propto
  \sqrt{
        \frac{1}{\omega -\Delta}
        \ln \left({\frac{1}{\omega -\Delta}}\right) 
         }
\quad\mbox{for}~0<\omega-\Delta<T_{K2}.
\label{log-correction}
\end{eqnarray}
%%%%%%%%%%%%%%%%%%%%%%%%%%%%%%%%%%%%%%%%%%%%%%%%%%%%%%%
The logarithmic correction appears due to the marginally irrelevant
ferromagnetic coupling.

In summary, we have studied the edge singularity in 
the optically induced  Kondo effect for a quantum dot 
with an even number of electrons.
We have shown that the Kondo exchange couplings 
generated in photo-excited states can be either antiferromagnetic or
ferromagnetic, which may lead to different anomalous behaviors
in the optical absorption spectrum according to the microscopic
parameters for a quantum dot.
It is interesting to observe 
such edge singularity in the absorption spectrum experimentally
by properly tuning the physical parameters such as the gate 
voltage and the magnetic field.

This work was partly supported by a Grant-in-Aid from the Ministry
of Education, Science, Sports and Culture, Japan.

%%%%%%%%%%%%%%%%%%%%%%%%%%%%%%%%%%%%%%%%%%%%%%%%%%%%%%%

%

\end{multicols}

\end{document}